# Telepath: Understanding Users from a Human Vision Perspective in Large-Scale Recommender Systems

**Yu Wang, Jixing Xu, Aohan Wu, Mantian Li, Yang He, Jinghe Hu, Weipeng P. Yan**

Business Growth Division. JD.com.
Beijing, China
{wangyu5, xujixing, wuaohan1, limantian, landy, hujinghe, paul.yan}@jd.com

**Abstract**

Designing an e-commerce recommender system that serves hundreds of millions of active users is a daunting challenge. From a human vision perspective, there're two key factors that affect users' behaviors: items' attractiveness and their matching degree with users' interests. This paper proposes Telepath, a vision-based bionic recommender system model, which understands users from such perspective. Telepath is a combination of a convolutional neural network (CNN), a recurrent neural network (RNN) and deep neural networks (DNNs). Its CNN subnetwork simulates the human vision system to extract key visual signals of items' attractiveness and generate corresponding activations. Its RNN and DNN subnetworks simulate cerebral cortex to understand users' interest based on the activations generated from browsed items. In practice, the Telepath model has been launched to JD's recommender system and advertising system. For one of the major item recommendation blocks on the JD app, click-through rate (CTR), gross merchandise value (GMV) and orders have increased 1.59%, 8.16% and 8.71% respectively. For several major ads publishers of JD demand-side platform, CTR, GMV and return on investment have increased 6.58%, 61.72% and 65.57% respectively by the first launch, and further increased 2.95%, 41.75% and 41.37% respectively by the second launch.

## Introduction

As one of the most popular online retailers in China, JD serves hundreds of millions of active users, displays a huge number of items and attains 100 billion dollars of GMV (gross merchandise value) yearly. However, this large scale of users and items obey very fat-tailed distributions. It brings great challenges in designing the recommender system, especially when facing billions of query requests and ranking tasks generated by users daily. These visits bring huge amounts of click-through log data daily, all weakly labeled. It indicates that these data can only provide a weak signal for an effective ranking model. In this case, there is an urgent need for an efficient way to model user behaviors. Through lots of statistics and analysis of the behaviors of JD users, we find two key factors that affect user's clicks and purchases. The first is the attractiveness of the content. Many outstanding creative items are often clicked even if the user has never browsed similar items. From this we can hypothesize that the creative content of the items bring new interests to user, because the user is vision oriented and usually has a visual curiosity. The other factor is whether the item matches the user's interests or purchase intentions. The two factors give us the directions to model user behaviors.

However, in a typical scene of e-commerce, all signals (appearances and descriptions of items) the user receives are captured only through the eyes, and the items browsed by the user directly show her potential interest. The above factors require a recommender system to be able to accurately extract the key visual signals that activate user interest or purchase intention and fully understand user interests.

(Brewer et al. 1998; Çukur et al. 2013) found that humans are good at finding contents of interest from complex, colored photographs, and the way that the visual signals captured from contents of interest attract humans is through activations in the cerebral cortex. The activations are processed and memorized by the cortex and affect functional areas of brain, finally forming conscious or subconscious interest. (Galli and Gorn 2011; Watanabe and Haruno) showed that both consciousness and subconsciousness affects human behaviors. (Silver et al. 2016; Taigman et al. 2014) showed that deep learning can reach human level performance in certain domains of application.

The above research and observations inspired us to build a ranking engine that extracts visual information like a human and is intelligent enough to understand users from a human vision perspective, thus addressing the aforementioned factors that affect user behaviors. Such a ranking engine needs two modules: a vision extraction module and an interest understanding module. The former simulates the human vision system to extract key visual signals from displayed items and generate vision activations; the latter simulates the human cerebral cortex to understand and memorize the user's conscious/subconscious interests based on the activations stimulated by the visual contents of the browsed items. In addition, the ranking engine also needs a scoring module that functions like a decision-making system to compute the match quality between the displayed item and the user's interest.

Driven by the success of deep convolution networks (CNNs) in computer vision (Szegedy et al. 2015, 2016, 2017) and recurrent neural networks (RNNs) in sequence process (Sutskever, Vinyals and Le 2014; Bahdanau, Cho, and Bengio 2014), we implemented the vision extraction module with CNN and the interest understanding module with RNN and DNN (deep neural network) respectively.

The CNN subnetwork generates activations from visual contents as would a human vision system. The activations are actually fixed-sized vectors and each represents an item whether or not it appears in training data. In addition to carrying rich vision information to increase expression ability, the module can also address the cold start problem caused by the fat-tailed distributions of items. The RNN subnetwork works to understand the user's conscious or immediate interest after processing the entire browsing sequence. As a complement, we apply a DNN subnetwork to understand the user's subconscious or comprehensive interest and personal preference by cross product transformations. The scoring module is a DNN subnetwork too.

Consequently, we propose Telepath, a vision based intelligent ranking paradigm. The primary contributions are as follows:

First, we find two key factors that affect user's clicks and purchases: attractive content of item and match with the user's interests. From a human vision perspective, we propose to simulate the user's vision system to extract the key signals that attract the user and generate vision activations, and simulate the cortex to understand the user's interests based on activations extracted from browsed items.

Secondly, we propose to apply a CNN subnetwork to simulate the user's vision system. In fact, this is a vision-based embedding method, which outperforms traditional embedding methods in terms of addressing the cold start problem and increasing expression power.

Lastly, we propose to apply RNN and DNN subnetworks to simulate the cerebral cortex. We apply RNN to understand the user's conscious interest; we apply DNN to understand the user's subconscious interest and preference.

In practice, Telepath has been launched to JD's production environment. For a major item recommendation block in JD app, the CTR (click-through rate), GMV and orders increased 1.59%, 8.16% and 8.71% respectively. For several major ads publishers of JD DSP (demand-side platform) (Wang et al. 2017), the CTR, GMV and ROI (return on investment) increased 6.58%, 61.72% and 65.57% respectively by the first launch, and further increased 2.95%, 41.75% and 41.37% respectively by the second launch.

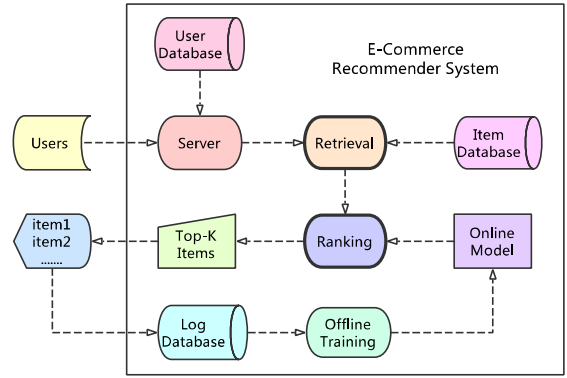

*Figure 1: Recommender System Flowchart*

## Architecture Design

JD's recommender system is illustrated in Figure 1. There are two key modules: the retrieval module and the ranking module. When a user visits the JD website or app, the recommender server is triggered and the retrieval module is requested to fetch candidate items. Then, the ranking module is called to predict a ranking score for each candidate. Finally, the top-K items are selected to display to the user. Ranking is the focus of this paper, but Telepath can also be applied in the retrieval module to trigger more items that are suitable. As an extension, we will discuss it in future work. Figure 2 gives a flowchart of Telepath that is composed of three modules: vision extraction, interest understanding and scoring.

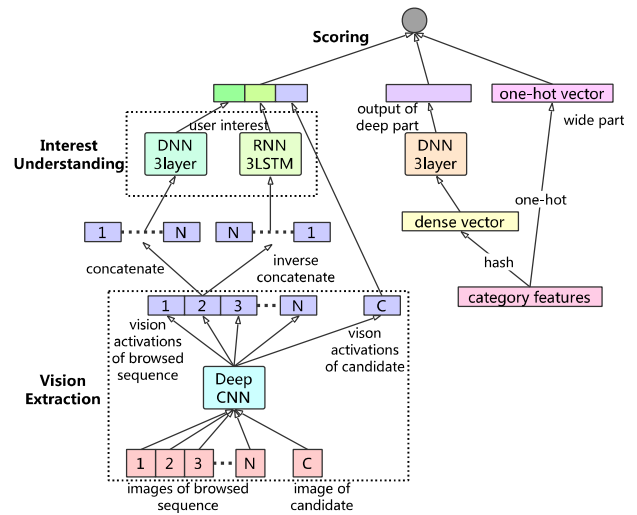

*Figure 2: Architecture of Telepath*

**Input data**

The visual contents that the user captures are images and description texts of displayed items. Images are usually more attractive to user, so we focus on images in this paper. However, the proposed method can also be used to process text. We will further explore this issue in our future work. The inputs for the vision extraction module are composed of $N$ latest browsed images and the image of candidate items. In order to speed up training, we scale each image to a size of 100×100. Each scaled image is converted into an RGB 3 channels matrix. The visual contents are captured from the user's vision perspective. However, there are other useful information that can help to improve accuracy. One is demographics, such as age, gender, geo location, user profile tags, etc., which are the prior information about user. The other is information about the item itself, such item id, category id, etc., which is the prior information about the item. These two types of category features can help to obtain statistics about users and items, for example, different preferences between men and women (or young and old), popular category (or item) in different geo locations (or ages), etc.

**Vision Extraction Module**

The vision extraction module is constructed to simulate the user's vision system to extract the key signals that attract the user to the displayed items and then generate vision activations that are represented as a dense vector for subsequent processing.

| Layer | Kernel Size | Input Size |
|---|---|---|
| Conv | 3 × 3/2 | 100 × 100 × 3 |
| Conv | 3 × 3/1 | 49 × 49 × 16 |
| Conv | 3 × 3/1, padding | 47 × 47 × 16 |
| MaxPool | 3 × 3/2, padding | 47 × 47 × 32 |
| Conv | 1 × 1/1 | 24 × 24 × 32 |
| 3×Block1 | N/A | 24 × 24 × 48 |
| 5×Block2 | N/A | 24 × 24 × 72 |
| 3×Block3 | N/A | 12 × 12 × 19 |
| AveragePool | 6 × 6/1 | 6 × 6 × 512 |
| FullConnection | 50 | 1 × 1 × 512 |

*Table 1: Architecture of the CNN subnetwork*

Traditional item-to-vector methods have two main shortcomings: the cold start problem and limited expression ability. These methods (Cheng et al. 2016; Covington et al. 2016) map each item to a vector through the common way, which is assigning each item an initialized dense vector and then correcting these vectors with large scale training data by a certain optimization method, e.g. stochastic gradient descent. It is undeniable that these methods can get high-quality vector representations for high-frequency items. However, there is a cold start problem. For low-frequency items and new items, it is difficult to get high-quality representations because these items do not get enough chances to be corrected. Even worse, this problem severely degrades the recommendation performance in large-scale e-commerce because most items in such domains are low-frequency, i.e. obeying fat-tailed distributions. In addition, the user is visually oriented and may be attracted by the whole content or only one tiny point. For example, a user wants to buy a new skirt and there are many very similar ones but with different colors. The user may only prefer a certain color, so only skirts of that color will be browsed. In this case, the color is the key signal that activates the user's click. Unfortunately, traditional item-to-vector methods cannot clearly express the color signal, which limits the expression ability.

Telepath addresses the above shortcomings from a human vision perspective. All the key signals that attract the user are contained in the visual contents that the user captures. And through a large number of different convolution kernels, CNN is good at extracting all kinds of visual signals. So we apply a CNN to simulate the user's vision system to extract the key visual signals. Finally, the captured signals are processed to generate a dense vector of output. From the perspective of the human vision system, the dense vector can be considered as the activations generated in the cerebral cortex, which carry rich visual information captured from raw inputs and can be further used to understand the user's interest. With a well-designed CNN architecture, the expression ability of activations can be greatly improved. Fortunately, driven by the success of work (Szegedy et al. 2016), we can design a better CNN architecture to process visual inputs as close as possible to a human vision system. The detailed architecture is shown in Table 1. Our Blocks 1-3 are similar to the Figures 5-7 in (Szegedy et al. 2016), but we apply a kernel of 5×5 instead of the original kernel of 3×3 and an average pooling of 3×3. The final full connection layer accepts an input of 1×1×512 and outputs a 50-dimensional vector which represents the vision activations used in the interest understanding module and the scoring module.

Note that the vision extraction module can generate activations from any displayed item no matter whether it appears in training data, thus addressing the cold start problem. In this paper, while CNN is the way we proposed to simulate the user's vision system, any well-designed CNN can be used to replace our proposed one.

**Interest Understanding Module**

As mentioned in the introduction, visual signals attract humans by generating activations in the cortex, and then these activations are understood and memorized by the cortex, finally forming interests. The browsed items are the ones that attracted the user in the recent past, which is feedback about the user's potential interest. So, based on the activations captured from the visual contents of the browsed items, the interest understanding module is able to simulate the cortex to understand users from a human vision perspective. To this end, there are two major tasks: understanding user's interest (i.e. conscious interest and subconscious interest) and extracting user's personal preferences (e.g. color, pattern, material, etc.). For example, if the user's most browsed items are red skirts, then we can infer that the user's purchase interest is 'skirt' and color preference is red. Telepath applies two subnetworks, RNN and DNN, to simulate the cortex for addressing the above tasks. RNNs are good at analyzing sequences and have human-level performance in language translation tasks; we hope RNNs can also succeed in understanding item browsing sequence. The browsed items form an ordered sequence and have potential dependency on each other because user's purchase intention varies gradually while browsing items. Through processing items one by one and storing accumulated information in memory cells, RNN can find the potential dependency, and then understand and simulate the changes of interest. Finally, RNN can represent the user's conscious interest. As shown in Figures 5, we find that feeding an inverse sequence to RNN can get a better performance than a forward sequence. We inferred that the reason for this is that the latest browsed items are more relative to the user's current interest. When it is first fed to RNN, the memory cells can get the information about the user's current interest earlier. Then it makes it easy for the optimization algorithm to find the optimum. A similar discovery can be found in language translation tasks (Sutskever, Vinyals, and Le 2014). DNN takes into account the whole sequence at the same time, so it can capture all the interests of the user that hide in the sequence, which can help to understand the user's comprehensive or subconscious interest. In addition, DNN is good at cross product transformations, which can help to find some commonalities that hide in the sequence and then further extract user's personal preferences.

We apply a 3-layer DNN and an RNN composed of a stack of $N = 3$ LSTM layers. The input of the RNN is the inverse browse sequence. As a complement, the input of the DNN is the forward sequence. We use the last output of RNN as user's conscious interest and the last output of DNN as user's subconscious interest. Both interest vectors are 50-dimension. The final user interests are composed of conscious and subconscious interests. If the user browses items with a certain interest and preference, this information will be captured and carried by the final interest vectors.

### Scoring Module

The scoring module takes the role of decision-making which is used to simulate the decision function area of the cortex to predict an item's score for final ranking based on the captured activations from the displayed candidate item, the extracted interest vectors of user and the additional prior information about user and item. An intuitive exploration is computing the matching quality between the user's interest vectors and the item's vision activations. The interest vectors and vision activations are both based on the visual contents that the user captures, which can be considered to understand the user from her vision perspective. In addition, in a typical e-commerce scenario, active users bring huge amounts of click-through log data daily. With this large amount of log data, the scoring module can get very useful statistics from indicators and their cross product transformations. Therefore, we apply wide and deep subnetworks to learn the statistics. For the wide part, each category feature is encoded to a one-hot vector. For the deep part, each category feature is mapped to a dense vector through a hash mechanism, and then these vectors are fed to another 3-layer DNN. Finally, these dense vectors will be corrected by an optimization algorithm based on training data. In fact, the wide part helps to learn memorization of feature interactions and the deep part helps to generalize better for unseen feature combinations. As shown in Figure 2, all extracted information are connected to a sigmoid layer to output the final ranking score.

## Experiments

In this section, we first discuss the visualization of vision activations of item and interest vectors of the users. Then, we discuss some offline experiment results. Lastly, we give the online experimental results obtained in two of JD's main recommendation application scenarios through A/B tests. The A/B test platform is similar to (Tang et al. 2010).

First, we give some hyperparameter configurations. In the CNN subnetwork, a dropout rate of 0.5 is applied in all layers to avoid over-fitting. The momentum for the updating gradient is set to 0.9 and the weight decay is set to 0.0005 for regularization. We employ a global learning rate of 0.01 for all layers that decays by a factor of 0.1 after every 100K steps with a mini-batch size of 64. Instead of non-synchronous ensemble training, we adopt joint training by back propagating the gradients from the output to all parts of the model simultaneously. In this paper, we use mini-batch stochastic optimization. We also use FTRL with L1 regularization to optimize the wide part and AdaGrad to optimize the other deep subnetworks.

### Visualization

The vision activations of an item are the output of the vision extraction module. For our item visualization experiment, we randomly selected six categories: watch, cellphone, down jacket, beach shoes, milk powder and cookies.

For each category, we randomly selected about 10,000 items. Figure 3 shows the visualization of the vision activations through t-SNE, in which the points with same color correspond to the same category. We can clearly see the clustering property of each category. However, we can also find a cross area between milk powder and cookies although they still have clear clustering trends. The reasons we infer are as follows. First, through jointly training, the Telepath model is trained to fit the user's clicks rather than a multi-class classification task. Second, there are many items in e-commerce that belong to different categories but have similar background colors or texture. Last, the vision extraction module tends to extract information that attracts users and matches user interest and personal preference, not just extract the category information. In this case, some captured activations of different categories of items may be projected to neighboring locations. For verifying our inference and giving an intuitive explanation, we show two pairs of examples in Figure 3, which are projected to neighboring locations. In each pair, the bottom one belongs to cookies and the other one belongs to milk powder. The left pair have similar cartoon patterns and the right pair have similar colors and baby patterns.

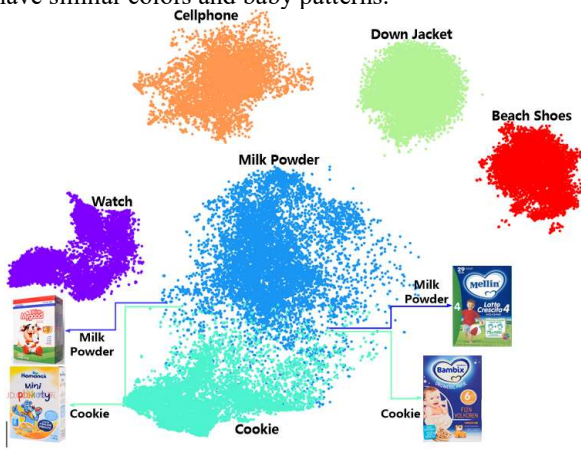

*Figure 3: Visualization of Vision Activations of Items*

In general, human's purchase interests are a mixture of conscious and subconscious, which are complex, diverse, and sometimes too subtle to be clearly distinguished. Typically, the more choices one has, the harder it is to make a decision.

For a certain degree of analysis, we assume that the item lastly browsed by a user represents her interest in the sense that Telepath should recommend that very item if she had not seen it yet. Based on this assumption, we select three categories of users who are interested in cookies, beach shoes and cellphones respectively. For each category, we randomly select about 6000 users. We generate interest activations for each user based on her recently browsed items other than the category item (i.e. the last browsed item). Figure 4(a) shows a visualization of all three categories of users. As expected, the activations are too subtle for t-SNE to do clear clustering. When considering only two categories, t-SNE tends to distinguish better as shown in Figure 4(c-d). This resembles a real-life phenomenon that having fewer choices always helps to make a decision.

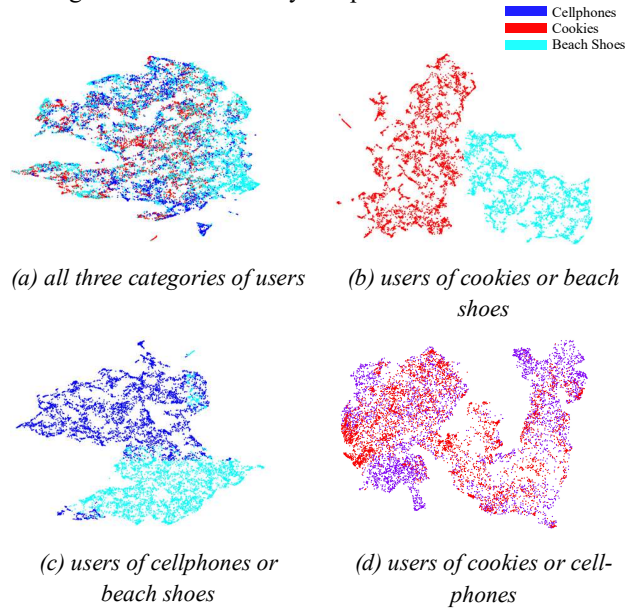

*(a) all three categories of users*     *(b) users of cookies or beach shoes*

*(c) users of cellphones or beach shoes*     *(d) users of cookies or cellphones*

*Figures 4: Visualization of User Interests Activations*

### Offline Experiments

We first discuss some offline experiments. In this section, all experiments are performed 5 times on training data and validation data, and finally we take the average performance. The experiments are divided into two groups. In the first group, the baseline is the traditional item-to-vector method implemented by Wide & Deep learning (Cheng et al. 2016) and Telepath only uses vision extraction module and interest understanding module (i.e. the left part of Figure 2). This group is to validate that the proposed vision extraction can outperform traditional item-to-vector and the proposed interest understanding is more effective than cross product transformations of DNN. To this end, the baseline only uses ids of items as features, and Telepath only uses images of items as features. In order to validate the better performance of feeding an inverse sequence to RNN, we also added an experiment that feeds a forward sequence to RNN as the additional baseline. In the second group, all mentioned features are used. Telepath is as shown in Figure 2. The baseline is Wide & Deep learning because the similar architecture has been launched in JD's online system. This group is to validate that Telepath can outperform a full featured Wide & Deep. For both groups, we analyze from two common evaluation metrics: log loss and AUC.

For the first group of experiments, Figure 5 shows that inversing the input sequence of RNN can get a better performance that verifies our previous analysis. Compared to the baseline, Telepath reduces the convergence value of loss from 0.675 to 0.665 and improves AUC from 0.664 to 0.693. The convergence speed of the baseline is faster than Telepath, but its final convergence value is higher than Telepath's. It can be inferred that vision embedding can extract richer information from visual inputs than traditional embedding, so it needs more time to explore more latent solution space, but brings a better performance. In addition, we can also infer that interest understanding is a more effective way to handle personal recommendations from the lower loss and the better AUC.

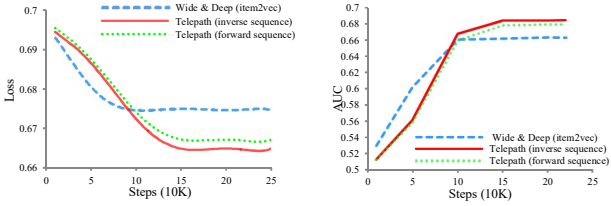

*Figures 5: ID Embedding versus Telepath Image Embedding*

For the second group of experiments, Figures 6 shows the joint performance. Compared to the baseline of Wide & Deep learning, Telepath reduces the convergence value of loss from 0.642 to 0.627 and improves AUC from 0.701 to 0.718. These results show the significant performance improvements of Telepath.

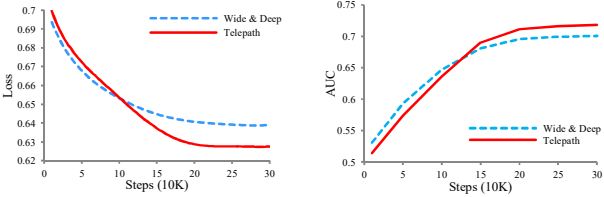

*Figures 6: Full-Featured Wide & Deep versus Telepath*

A comparison of the right-hand graphs of Figure 5 and Figures 6 show that even only using images, Telepath can reach an AUC that is comparable to a full-featured Wide & Deep with billions of weights, which shows the high efficiency of vision extraction and interest understanding.

**Online Experiments**

The Telepath model has been launched on two main recommendation application scenarios within JD. The first one is item recommendation block X in the JD app. The block X is very important to JD because it is visited billions of times every day and JD needs to finish billions of recommendations correspondingly. With better recommendations, trading orders can be significantly increased. For item recommendation business, we pay more attention to the improvements on CTR, trading orders and GMV.

The second one is online advertising in publishers Y of JD DSP. Publishers Y are several of JD's main partners in DSP business. The daily-recommended ads that JD makes for users that are visiting publishers Y also reach the scale of hundreds of millions. With good recommendations, more users will be driven to the JD website or app. For advertising recommendation business, we pay more attention to the improvements on CTR, GMV and ROI (GMV/revenue). For the first two online experiments, the baseline is JD's previous ranking model that has a structure similar to Wide & Deep learning with billions of weights.

| Date | Day1 | Day2 | Day3 | Day4 |
|---|---|---|---|---|
| CTR | +0.02% | +2.37% | +1.93% | +2.84% |
| GMV | +15.04% | +7.81% | -2.36% | +10.05% |
| Orders | +6.62% | +5.10% | +8.54% | +13.92% |
| Date | Day5 | Day6 | Day7 | Average |
| CTR | +0.62% | +2.36% | +0.97% | **+1.59%** |
| GMV | +6.77% | +8.36% | +11.48% | **+8.16%** |
| Orders | +9.90% | +12.17% | +4.7% | **+8.71%** |

*Table 2: Improvements on Item Recommendation Block X*

| Date | Day1 | Day2 | Day3 | Day4 |
|---|---|---|---|---|
| CTR | +5.15% | +8.07% | +10.5% | +6.15% |
| GMV | +126.48% | +9.1% | +18.4% | -19.24% |
| ROI | +129.53% | +14.35% | +14.2% | -17.44% |
| Date | Day5 | Day6 | Day7 | Average |
| CTR | +4.63% | +2.11% | +9.48% | **+6.58%** |
| GMV | +8.53% | +143.09% | +145.74% | **+61.72%** |
| ROI | +9.17% | +161.36% | +147.79% | **+65.57%** |

*Table 3: Improvements on Publishers Y (Launched Version 1)*

| Date | Day1 | Day2 | Day3 | Day4 |
|---|---|---|---|---|
| CTR | -1.56% | +14.76% | -2.57% | +0.85% |
| GMV | +81.94% | +108.59% | +24.29% | +22.81% |
| ROI | +83.61% | +106.83% | +19.48% | +23.3% |
| Date | Day5 | Day6 | Day7 | Average |
| CTR | +0.34% | +8.40% | +0.42% | **+2.95%** |
| GMV | -33.50% | +30.18% | +57.94% | **+41.75%** |
| ROI | -35.05% | +33.56% | +57.84% | **+41.37%** |

*Table 4: Improvements on Publishers Y (Launched Version 2)*

Table 2 shows the online performance of item recommendation block X in the JD app. On average, CTR, GMV and orders have increased 1.59%, 8.16% and 8.71% respectively. Performance gains that are greater than 1.0% over 7 days are considered statistically significant in the real recommender system. Table 3 shows the performance of JD's advertising recommendations when first launching online. On average, CTR, GMV and ROI have increased 6.58%, 61.72% and 65.57% respectively. For online advertising in publishers Y of JD DSP, there are two display styles for ads: one image or three images. For ads with three images, the first launched version of Telepath only uses the first image. However, in practice, ads with three images are more attractive. Therefore, we launched the second version of Telepath. For ads with one image, the single image is used. For ads with three images, all three images are used. Table 4 shows the performance of the second launched version. Baseline is the first launched version. On average, CTR, GMV and ROI have further increased 2.95%, 41.75% and 41.37% respectively.

The excellent performance of Telepath on JD's item and advertising recommendations, especially the significantly improved GMV, orders and ROI, indicate that recommendations through Telepath are more likely to be browsed and further purchased. This proves that Telepath can accurately extract the key visual signals that attract users through the vision extraction module and fully understand user's interests through the interest understanding module, thus actually addressing the aforementioned factors that affect user's behaviors.

## Related Work

There have been many literatures proposed for recommender systems (Davidson et al. 2010; Rendle, Freudenthaler and Schmidt-Thieme 2010; Lops, De Gemmis and Semeraro 2011) to which readers should refer for in-depth details. As the key module in recommender systems, ranking strategy is the key to improving recommending performance, which is often based on learning-to-rank (Cao et al. 2007; Qin et al. 2008; Liu 2009) that emerges from late 1990s.

Driven by the success of deep learning, DNNs have been applied to large-scale recommender systems. One of the popular works is the video recommendations on YouTube (Covington, Adams, and Sargin 2016) which applies deep learning for both candidate generation and ranking. In the work, each video is mapped to a separate learned embedding and out-of-vocabulary videos are simply mapped to a zero embedding. Another popular work is the app recommendations on Google Play (Cheng et al. 2016) whose key idea is to learn memorization through a wide set of cross-product feature transformations and generalizations for unseen feature combinations through DNNs. Similarly, each unique app is mapped to an embedding vector that is initialized randomly and is corrected by training. Unfortunately, neither of the above works gave an effective way to resolve the cold start problem.

In language models, RNNs have achieved great success, especially for sequence to sequence tasks, which has inspired researchers to apply RNNs in recommender systems (Wu et al. 2016; Ko, Maystre and Grossglauser 2016). In computer vision, CNNs have reached high accuracy in image classification and object recognition tasks (Krizhevsky, Sutskever, and Hinton 2012; Szegedy et al. 2016). Deep residual networks (He et al. 2016) have been used to reduce the difficulty of training deeper models through skipping one or more layers. This makes it possible to train deeper and wider networks (Szegedy et al. 2017).

## Future Work

For the visual contents captured by the user, this paper focuses on items' images, but the description text is also important visual content from which the user can obtain additional information like size, weight, etc. Therefore, one future goal is to add a text-understanding subnetwork to the vision extraction module. To simulate humans understanding text and extracting key text signals, we plan to apply a CNN network similar to Table 1 to process the character level one-hot encoded matrix of the text. In addition to the description texts of displayed items, we also plan to take into account the user's historical query texts to better understand their interest.

The key method in this paper is to understand the user from a human vision perspective. This method can also be applied to the retrieval module. The retrieval module is to trigger more candidates that attract the user to increase the competitions in the ranking stage. The proposed method can help to trigger more suitable items from the user's vision perspective.

## Conclusion

This paper proposes Telepath, a vision-based intelligent ranking paradigm that understands users from a human vision perspective to address the aforementioned two key factors that affect user behaviors. Telepath use a CNN to simulate the user's vision system to extract the key signals that attract the user and generate vision activations, and a combination of RNN and DNN to simulate the cerebral cortex to understand and memorize user interests and preferences.

To the best of our knowledge, such a ranking paradigm has not been proposed in previous works. In practice, Telepath has achieved significant improvements with regard to online recommendations and ads business for JD, which proves its ability of better understanding users.